\documentclass{iaus}
\usepackage{graphicx}

\title[IAU 253.~~Transiting planets] %% give here short title %%
{Measuring accurate transit parameters}

\author[Joshua N.\ Winn]  %% give here short author list %%
{Joshua N.\ Winn}

\affiliation{Department of Physics, and Kavli Institute
  for Astrophysics \& Space Research \\
  Massachusetts Institute of Technology \\
  77 Massachusetts Avenue \\
  Cambridge, MA 02139-4307 \\
  email: {\tt jwinn@mit.edu}}

\pubyear{2008}
\volume{253}  %% insert here IAU Symposium No.
\pagerange{XXX--YYY}
\jname{Transiting Planets}
\editors{eds.~F.~Pont et al.}
\begin{document}

\maketitle

\begin{abstract}
  By observing the transits of exoplanets, one may determine many
  fundamental system parameters. I review current techniques and
  results for the parameters that can be measured with the greatest
  precision, specifically, the transit times, the planetary mass and
  radius, and the projected spin-orbit angle.
\end{abstract}

\firstsection % if your document starts with a section,
\section{Introduction}

Henry Norris Russell~(1948) once delivered a lecture here in Cambridge
entitled ``The royal road of eclipses,'' about the determination of
accurate parameters for eclipsing binary stars, and the promise that
such systems held for progress in stellar astrophysics. Given the
rapid progress on display at this meeting, it is clear that
exoplanetary science too has its royal road: the royal road of
transits. Figure~1 illustrates the happy situation in which the
planet's orbit is viewed nearly edge-on, and the planet undergoes
transits and occultations.

\begin{figure}[b]
% \vspace*{-2.0 cm}
\begin{center}
 \includegraphics[width=29pc]{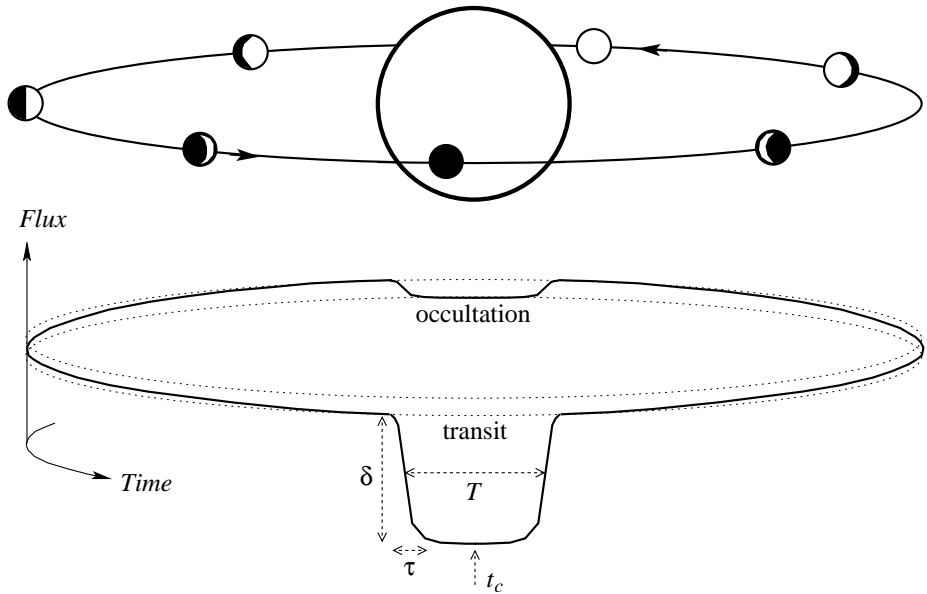}
% \vspace*{-1.0 cm}
 \caption{Illustration of transits and occultations. During a transit,
   the planet blocks a fraction of the starlight.  Afterwards, the
   planet's brighter dayside comes into view and the total flux
   rises. The total flux drops again when the planet is occulted by
   the star.}
   \label{fig1}
\end{center}
\end{figure}

\begin{table}
%  \begin{center}
  \caption{Properties that have been measured, or that might be
    measured in the future, through precise observations of transiting
    planets.}
    \label{tab1}
    \begin{tabular}{lclc}\hline
      {\bf Property} & {\bf Refs.} & ~~~~~~~{\bf Property} & {\bf Refs.} \\ \hline
      Orbital period & 1,2 & ~~~~~~~Planet-planet interactions (short-term) & 19,20 \\
      Orbital inclination & 1,2 & ~~~~~~~Planet-planet interactions (long-term) & 21,22 \\
      Planetary mass & 1,2 & ~~~~~~~Mutual orbital inclinations & 20,23 \\
      Planetary radius & 1,2 & ~~~~~~~Planetary rings & 24,25 \\
      Stellar obliquity & 3,4 & ~~~~~~~Satellites & 9,24 \\
      Orbital eccentricity & 5,6 & ~~~~~~~Relativistic precession & 26,27 \\
      Stellar limb darkening & 7 & ~~~~~~~Parallax effects & 28, 29 \\
      Star spots & 8,9 & ~~~~~~~Apsidal motion constant & 30 \\
      Thermal emission & 5,10 & ~~~~~~~Stellar differential rotation & 31 \\
      Absorption spectrum & 11,12 & ~~~~~~~Oblateness and obliquity & 32,33 \\
      Albedo & 13,14 &  ~~~~~~~Variations in stellar radius & 34 \\
      Phase function & 15 & ~~~~~~~Yarkovsky effect & 35 \\
      Effective radiative time constant & 16 & ~~~~~~~Planetary wind speed & 36 \\
      Trojan companions & 17,18 &  ~~~~~~~Artificial planet-sized objects & 37 \\
      \hline
    \end{tabular}

%  \end{center}
%  \vspace{1mm}
  {\it Non-exhaustive list of references:}
  (1) Charbonneau et al.~(2000). (2) Henry et al.~(2000).
  (3) Queloz et al.~(2000). (4) Winn et al.~(2005).
  (5) Charbonneau et al.~(2005). (6) Bakos et al.~(2007).
  (7) Knutson et al.~(2007a).
  (8) Silva~(2003). (9) Pont et al.~(2007).
  (10) Deming et al.~(2005).
  (11) Charbonneau et al.~(2002). (12) Vidal-Madjar et al.~(2003).
  (13) Rowe et al.~(2006). (14) Winn et al.~(2008a).
  (15) Knutson et al.~(2007b).
  (16) Langton \& Laughlin~(2008).
  (17) Ford \& Gaudi~(2006), (18) Madhusudhan \& Winn (2008).
  (19) Holman \& Murray~(2005). (20) Agol et al.~(2005).
  (21) Miralda-Escud\'e (2002). (22) Heyl \& Gladman~(2007).
  (23) Fabrycky, D., these proceedings.
  (24) Brown et al.~(2001). (25) Barnes \& Fortney~(2004).
  (26) P\'al \& Kocsis~(2008). (27) Jordan \& Bakos (2008).
  (28) Scharf~(2007). (29) Rafikov~(2008).
  (30) Ragozzine \& Wolf~(2008).
  (31) Gaudi \& Winn~(2007).
  (32) Seager \& Hui~(2002), (33) Barnes \& Fortney~(2003).
  (34) Loeb~(2008).
  (35) Fabrycky~(2008).
  (36) Spiegel et al.~(2007).
  (37) Arnold~(2005).
\end{table}

Table~1 summarizes the information that has been obtained---or that is
obtainable in principle---through precise observations of transits and
occultations. This table is surely incomplete. Every few months, a new
and creative application of transit observations is proposed. I was
asked to discuss some of the measurements that can be made with the
highest signal-to-noise ratio. In the best cases, we can measure
orbital periods with 8 significant digits; transit times to within a
fraction of a minute; the planetary mass and radius to within a few
per cent; and the stellar obliquity (or at least its sky projection)
to within a few degrees.

\section{Transit light curve parameters}

In any discussion of measuring accurate transit parameters, the first
question should be: what are those parameters? Ignoring limb darkening
for the moment, the 4 basic observables are (with reference to Fig.~1)
the mid-transit time $t_c$, the depth $\delta$, the total duration
$T$, and the partial duration $\tau$. These observables can be
translated into 3 dimensionless parameters describing the physical
properties of the system:
\begin{eqnarray}
  {\rm Radius~ratio}~& R_p/R_s & \approx~\sqrt{\delta}, \\
  {\rm Impact~parameter}~& b & \approx~1 - \sqrt{\delta}\frac{T}{\tau}, \\
  {\rm Scaled~stellar~radius}~& R_s/a &
  \approx~\frac{\pi\sqrt{T\tau}}{\delta^{1/4}P}
  \left( \frac{1+e\sin\omega}{\sqrt{1-e^2}} \right),
\end{eqnarray}
where $R_p$ and $R_s$ are the planetary and stellar radii; $b$ is the
impact parameter; $P$ is the orbital period; and $e$ and $\omega$ are
the orbital eccentricity and argument of pericenter, which can be
measured from the Doppler data. The approximations given here are
valid for small values of $\delta$, $\tau/T$, and $T/P$, and they
neglect limb darkening. Some useful references are Seager \&
Mallen-Ornelas~(2003), who give the exact correspondences for a
circular orbit, and Carter et al.~(2008), who derive analytic
expressions for the errors in these quantities. Mandel \& Agol~(2002)
and Gimenez~(2006) provide codes for calculating realistic light
curves including limb darkening.

The dimensionless parameters of Eqns.~(2.1-2.3) are useful, but
oftentimes one is more interested in dimensionful parameters such as
the planetary mass in Jovian masses, or the semimajor axis in AU. For
those, the light curve analysis must be supplemented with some
external information, such as an estimate of the stellar mass and/or
radius based on the stellar parallax, spectrum, angular diameter, and
any other available data. Interestingly, though, if one is willing to
trust Kepler's Law, then the transit observables may be combined into
two quantities with dimensions (Seager \& Mallen-Ornelas 2003,
Southworth et al.~2007):
\begin{eqnarray}
{\rm Stellar~mean~density}~& \rho_s & \approx~\frac{3P}{\pi^2 G}
\left( \frac{\sqrt{\delta}}{T\tau} \right)^{3/2}
\left[ \frac{1-e^2}{(1+e\sin\omega)^2} \right]^{3/2} \\
{\rm Planetary~surface~gravity}~& g_p & \approx~\frac{2\pi K_s}{P}
\frac{\sqrt{1-e^2}}{\delta(R_s/a)^2 \sin i},
\end{eqnarray}
where $K_s$ is the velocity semi-amplitude of the stellar
radial-velocity signal, and $i$ is the orbital inclination. The
inclination can be written in terms of observables using
\begin{equation}
b = \frac{a\cos i}{R_s} \left( \frac{1-e^2}{1+e\sin\omega} \right).
\end{equation}
Knowledge of the stellar mean density is helpful for pinning down the
stellar properties. The photometrically-determined $\rho_s$ is
superior as a gravity indicator than the traditional $\log g$ that is
based on the widths of pressure-sensitive absorption lines (see, e.g.,
Sozzetti et al.~2007, Winn et al.~2008b). Recently, Torres et
al.~(2008) put this technique into practice for 23 transiting
exoplanets, in the most homogeneous and complete analysis of transit
data to date. A similar effort is underway by Southworth~(2008). The
relative immunity of $g_p$ to systematic errors in the stellar
properties suggests that when testing theoretical models of planetary
structure, it would be wiser to compare theoretical and observed
surface gravities, rather than the traditional comparison between
theoretical and observed radii.

\begin{figure}[t]
% \vspace*{-2.0 cm}
\begin{center}
 \includegraphics[width=32pc]{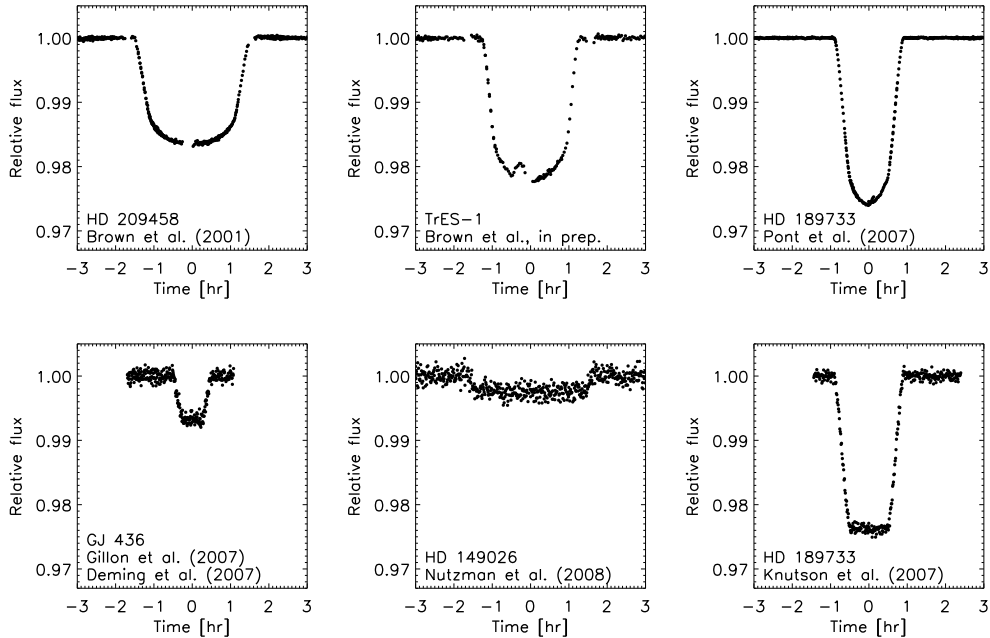}
% \vspace*{-1.0 cm}
 \caption{A gallery of transit light curves based on observations
with spaceborne instruments. The upper 3 panels show optical data
from the {\it Hubble Space Telescope}\, and the lower 3 panels
show infrared data from the {\it Spitzer Space Telescope}.}
   \label{fig2}
\end{center}
\end{figure}

This field has been blessed with spectacular data from the space
observatories, some of which are shown in Fig.~2. Everyone remembers
exactly where they were when they first saw the {\it Hubble
  Space Telescope}\, ({\it HST}\,) light curve of HD~209458 by Brown
et al.~(2001), with a cadence of 80~s and a precision of $1.1\times
10^{-4}$. This group established the protocol of most subsequent {\it
  HST}\, observations: schedule several ``visits'' to achieve complete
phase coverage of the event; disperse the starlight over many pixels
to average over sensitivity variations and achieve a high duty cycle;
sum all of the detected photons; and correct for instabilities due to
the spacecraft orbit and detector.

With the {\it Spitzer Space Telescope}\, the count rates are generally
lower, but there are compensatory advantages: there is little
limb-darkening at at mid-infrared wavelengths, and uninterrupted views
of entire events are possible because the satellite is not in a
low-earth orbit. A few noteworthy Spitzer results are the transit of
GJ~436 (Gillon et al.~2007, Deming et al.~2007), the Neptune-sized
planet about which we heard so much this week, and a recent light
curve by Nutzman et al.~(2008) of the very challenging target
HD~149026, for which the transit depth is only a quarter of a
percent.

Ground-based photometry still plays an important role. With a large
claim on a small telescope, one can observe many more events than is
feasible from space, build a large library of transit times, and
respond more quickly to new discoveries. Furthermore, in many cases it
has been possible with ground-based photometry to reach the regime in
which the contribution to the parameter uncertainties due to
photometric errors is comparable to the systematic errors arising from
uncertainties in the stellar properties. Beyond this point, improved
photometric precision gives steeply diminishing returns. The following
discussion is informed by my experience with the Transit Light Curve
project, which Matt Holman and I have undertaken (see Fig.~3). It
should be emphasized, however, that spectacular ground-based
photometry has been achieved by several other groups (see, e.g.,
Alonso et al.~2008, or Gillon,~M.\ in these proceedings).

\begin{figure}[t]
% \vspace*{-2.0 cm}
\begin{center}
 \includegraphics[width=32pc]{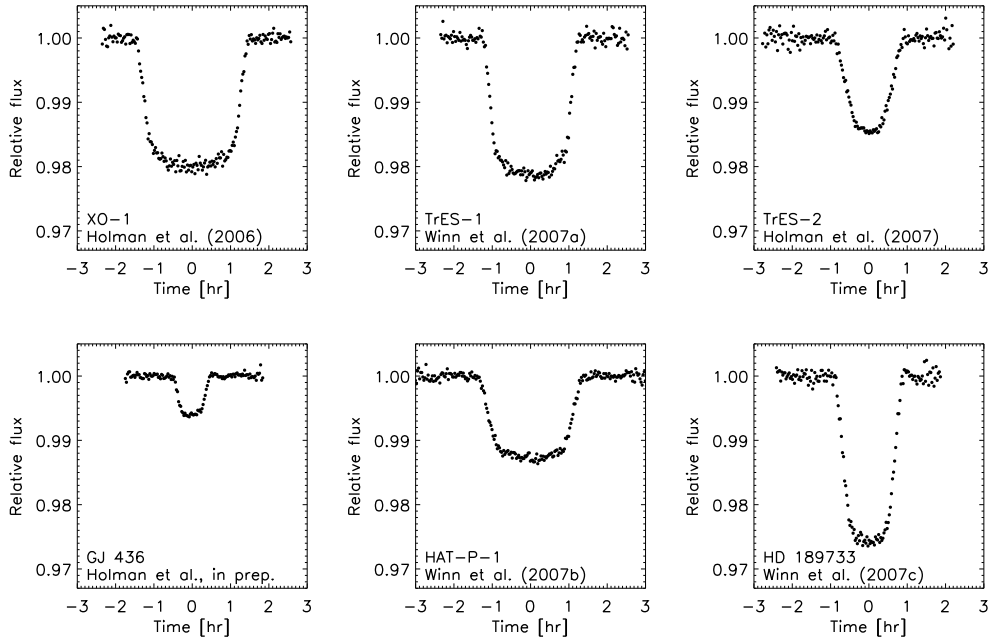}
% \vspace*{-1.0 cm}
 \caption{A gallery of transit light curves, based on observations
   with small ground-based telescopes. Each panel shows a
   time-averaged (100~s), composite light curve based on observations
   of multiple transits. The number of observed transits ranges from 2
   (XO-1) to 11 (GJ~436).}
   \label{fig3}
\end{center}
\end{figure}

The aspiring ground-based transit observer is well-advised to consult
some of the classic papers on CCD differential ensemble photometry,
including Gilliland \& Brown~(1988), Kjeldsen \& Frandsen~(1992), and
Everett \& Howell~(2001). Some lessons from these authors, as applied
to transit observations, are to ensure that calibration frames have
negligible photon-counting noise; keep the pointing stable to within a
few pixels to minimize the effects of pixel-to-pixel gain variations;
and strive to observe at least 10 comparison stars bracketing the
target star in brightness and color. Each instrument will have a sweet
spot, a magnitude range for which a 1~min exposure gives $\sim$10 good
comparison stars within the field of view. For TLC observations,
employing the 1.2m telescope at the Fred L.~Whipple Observatory and
the Keplercam $23'$ detector, this occurs at 11-12th magnitude. For
brighter targets, defocusing is helpful to draw out the maximum
exposure time (and thereby increase the duty cycle) and to hedge
against pixel-to-pixel variations and seeing variations.

It is often advisable to use a long-wavelength bandpass, where
extinction variations are smaller and the effects of stellar limb
darkening are reduced. Smaller limb darkening leads to light curves
with sharper corners and flatter bottoms, providing more statistical
leverage on the parameters $t_c$, $R_p/R_s$, $b$, and $R_s/a$. To be
concrete, Fig.~4 shows the results of calculations by P\'al~(2008),
comparing the statistical error in those parameters as a function of
the observing bandpass. All other things being equal, observing in $K$
as opposed to $u$ reduces the statistical error in $t_c$ by
$\sim$15\%, in $R_s/a$ or $b$ by $\sim$40\%, and $R_p/R_s$ by
$\sim$80\%. Of course, there are other factors that affect the choice
of bandpass, such as the count rate; the purpose of Fig.~4 is to
isolate the effect of limb darkening.

\begin{figure}[b]
\vspace*{-0.5 cm}
\begin{center}
 \includegraphics[width=32pc]{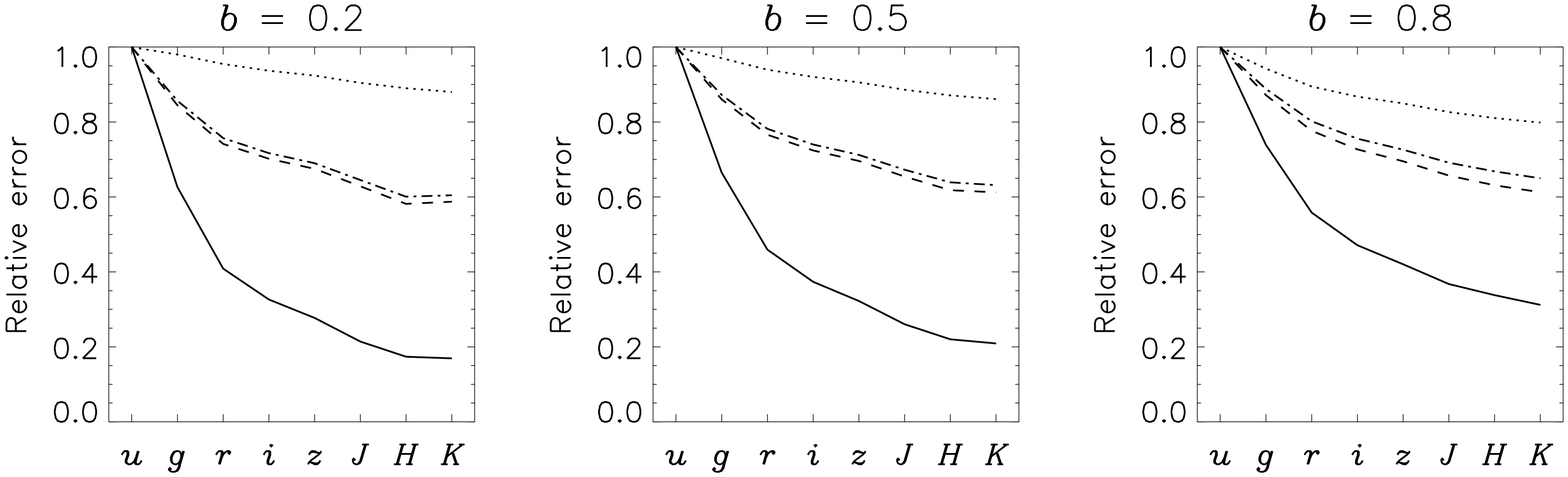}
\vspace*{-1.0 cm}
\caption{ The effect of limb-darkening on parameter errors, based on
  calculations by P\'al ~(2008). Imagine a planet with $R_p/R_s=R_s/a
  = 0.1$ transits a Sun-like star with impact parameter $b$, and a
  light curve is obtained with $1.3\times 10^{-3}$ precision and 10~s
  cadence. Shown here is the relative error in the parameters
  $R_p/R_s$ (solid), $b$ (dashed), $a/R_s$ (dash-dotted), and $t_c$
  (dotted), as a function of the observing bandpass. For redder
  bandpasses (to the right), the effect of limb-darkening is smaller
  and the parameter errors are decreased. }
   \label{fig4}
\end{center}
\end{figure}

A primary goal of high-precision transit observations is to
characterize the mass and radius of the planet. Figure~5 shows results
from the current ensemble. When possible, the results were taken from
the homogeneous analysis of Torres et al.~(2008) mentioned previously,
and otherwise from the most recent literature. Almost all of the
planets seem to be gas giants, with radii 10-50\% larger than that of
Jupiter. A persistent theme in this field is that at least a few
planets have radii that are ``too large'' by the standards of
theoretical models of H-He giant planets, even after accounting for
the intense stellar heating and selection effects (see Burrows et
al.~2007 for a recent discussion). In the left panel, with logarithmic
axes, one may appreciate that the Neptune-sized planet GJ~436 is
``halfway'' from the gas giants to Earthlike planets.

\begin{figure}[bh]
% \vspace*{-2.0 cm}
\begin{center}
 \includegraphics[width=32pc]{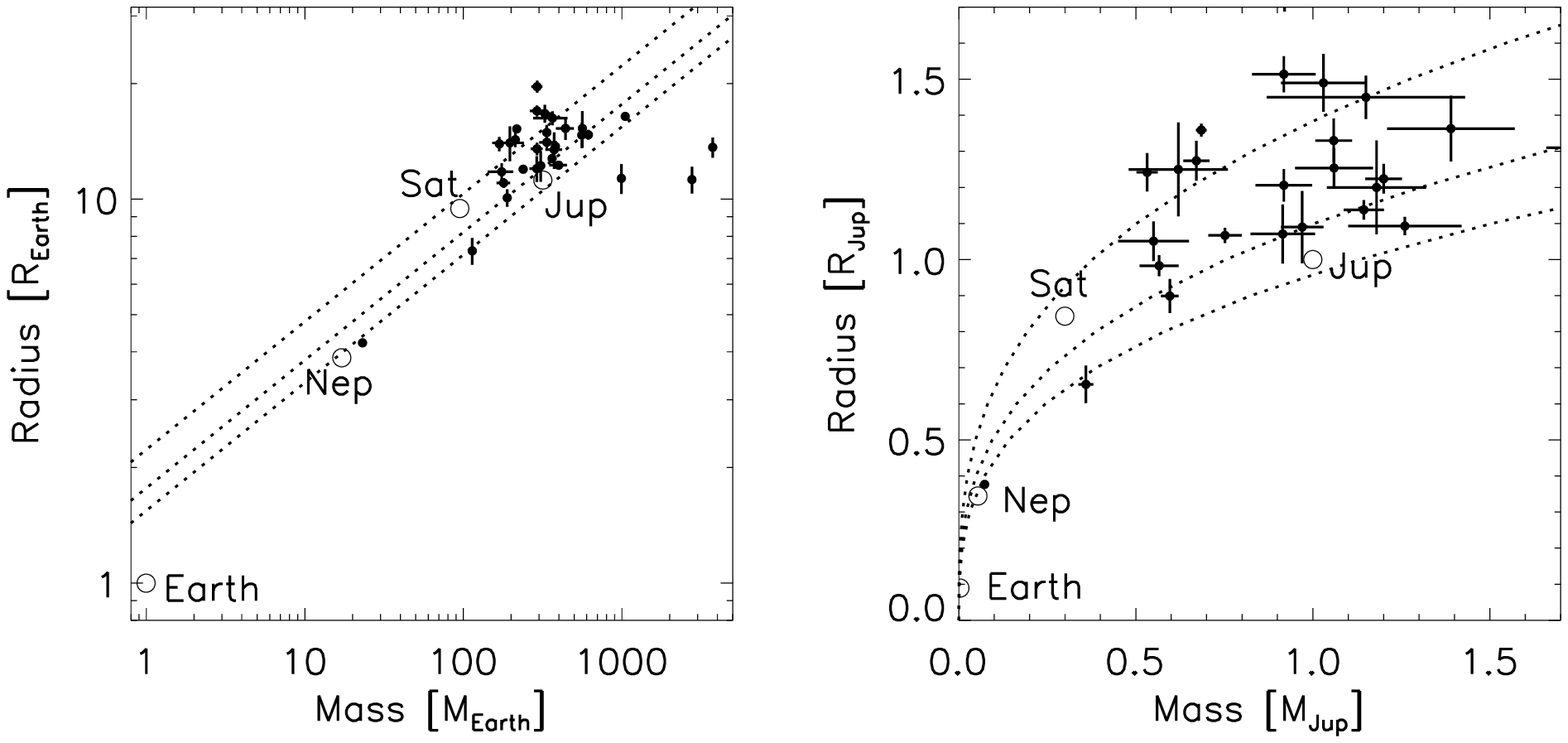}
% \vspace*{-1.0 cm}
\caption{Masses and radii of transiting planets. The dotted lines are
loci of constant mean density. Values for Jupiter, Saturn, Neptune,
and Earth are also plotted, for comparison.}
\label{fig5}
\end{center}
\end{figure}

Another reason to observe transits is to measure precise mid-transit
times. These data can be used to detect additional planets using the
method of Holman \& Murray~(2005) and Agol et al.~(2005):
planet-planet gravitational forces cause slight modifications in the
orbit of the transiting planet, which are revealed by a pattern of
anomalies in a sequence of transit times. This technique is especially
sensitive to planets in mean-motion resonance with the transiting
planet. The acquisition of transit times is being pursued by a large
number of groups, many of whom have presented results at this meeting.
Dynamical analyses by Agol \& Steffen~(2007), Miller-Ricci et
al.~(2008), and others have not yet resulted in any secure detections,
but in the best cases they have ruled out terrestrial-mass planets in
the lowest-order resonances.

As another illustration of the extreme sensitivity to resonant orbits,
consider GJ~436. Circumstantial evidence for a super-Earth in the 2:1
resonance with the transiting planet was presented by Ribas et
al.~(2008). Figure~6, produced by D.~Fabrycky, shows some recently
measured transit times for GJ~436 along with the timing variations
that one expects from a 5~$M_\oplus$ planet in the 2:1 resonance
(solid lines). The amplitude of the theoretical timing variations are
$\sim$$10^4$~s, far out of the plotting range, and are strongly
excluded by the data. The largest allowed body in that particular
orbit is about 0.05~$M_\oplus$, or 4 Lunar masses (dotted lines).

One may also search for planets in 1:1 resonance: Trojan planets,
residing at the L4/L5 points of the planet's orbit. Ford \&
Gaudi~(2006) showed that Trojans can be detected by seeking a
difference between the observed transit time and the calculated
transit time based only on the radial-velocity data. This technique
was recently applied to 25 transiting systems, and did not result in
any secure detections, although for GJ~436 it was possible to rule out
Trojans of mass 2.5~$M_\oplus$ or larger (Madhusudhan \& Winn~2008).

\begin{figure}[t]
% \vspace*{-2.0 cm}
\begin{center}
 \includegraphics[width=28pc]{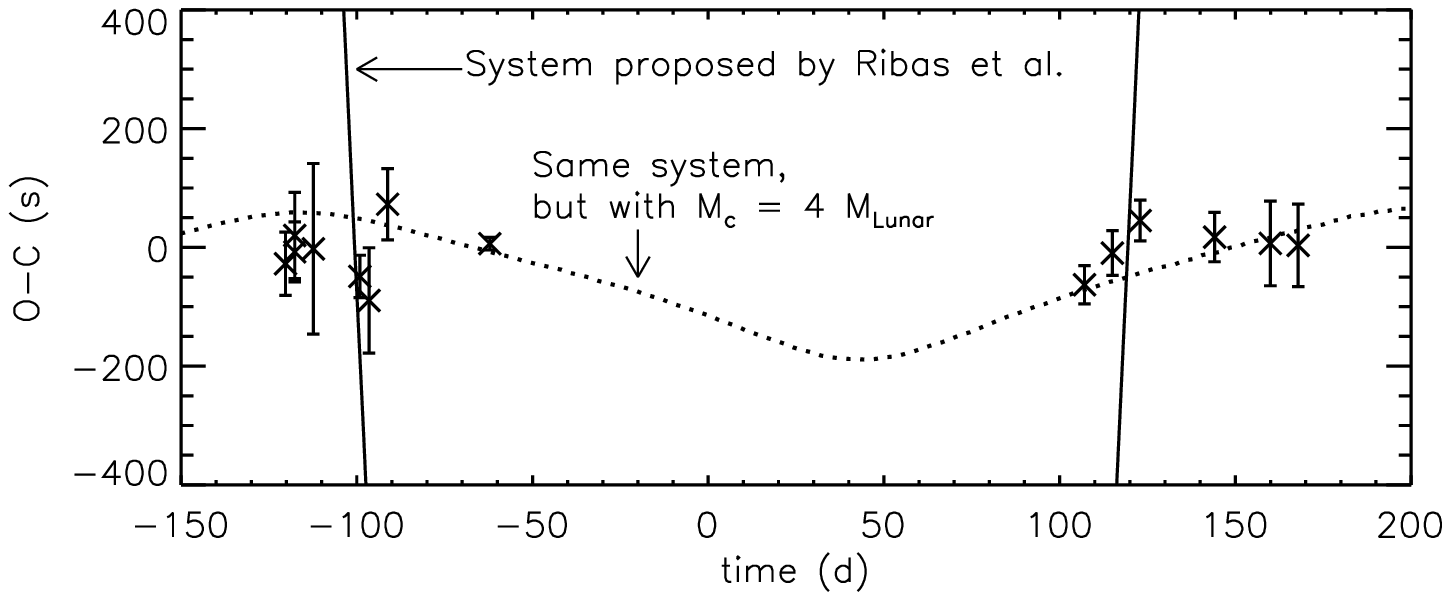}
% \vspace*{-1.0 cm}
 \caption{Transit timing residuals (observed~$-$~calculated) for
   GJ~436, along with theoretical variations that would be expected
   due to a second planet in a 2:1 resonance, with a mass of
   5~$M_\oplus$ (solid line) and 0.05~$M_\oplus$ (dotted line).
   Calculations and figure by D.~Fabrycky.}
\label{fig6}
\end{center}
\end{figure}
 
\section{The Rossiter-McLaughlin effect}

The last parameter I will discuss is the stellar obliquity, or
spin-orbit angle, defined as the angle between the angular momentum
vectors of the rotating star and of the planetary orbit. The
spin-orbit angle is a fundamental geometric property, and for this
reason alone, it is worth seeking empirical constraints on $\psi$
whenever possible. In addition, it has recently been recognized that
$\psi$ is a possible diagnostic of theories of planet migration.
Migration via tidal interactions with a protoplanetary disk would
probably result in a close spin-orbit alignment. Migration due to
planet-planet scattering events, or Kozai cycles accompanied by tidal
friction, would produce at least occasionally large misalignments
(see, e.g., Fabrycky \& Tremaine 2007, Wu et al.~2007, Chatterjee et
al.~2008, Nagasawa et al.~2008, Juric \& Tremaine 2008). Independently
of the interpretation, one may regard $\psi$ to be on a par with the
semimajor axis and the eccentricity: all are basic geometric
parameters, for which accurate and systematic measurements can lead to
revealing discoveries and statistical constraints on exoplanetary
system architectures.

Alas, the spin-orbit angle is not generally measurable. However, by
monitoring the apparent Doppler shift of the host star throughout a
transit (in addition to the loss of light) one may determine the angle
between the {\it sky projections}\, of the two angular momentum
vectors, which is denoted $\lambda$ and referred to as the projected
spin-orbit angle. It is then possible to derive a lower limit on
$\psi$ for an individual system, and to derive statistical constraints
on $\psi$ from an ensemble of results.

The sensitivity to $\lambda$ arises from a spectroscopic phenomenon
described by Rossiter (1924) and McLaughlin (1924) for the case of
eclipsing binaries, and known today as the RM effect. During a
transit, the planet blocks a portion of the rotating stellar surface,
thereby hiding some of the velocity components that ordinarily
contribute to the stellar line broadening. The result is a distorted
line profile that is usually manifested as an ``anomalous'' Doppler
shift. When the planet is projected in front of the approaching
(blueshifted) half of the star, the net starlight appears slightly
redshifted, and vice versa. Figure~7 shows three trajectories of a
transiting planet that have the same impact parameter---and hence
produce identical light curves---but that have different orientations
relative to the stellar spin axis, and hence produce different RM
signals. The signal for a well-aligned planet is antisymmetric about
the midtransit time (left panels), whereas a strongly misaligned
planet that spends all its time in front of the receding half of the
star will produce only an anomalous blueshift (right panels).

\begin{figure}[t]
 \vspace*{-0.75 cm}
\begin{center}
 \includegraphics[width=32pc]{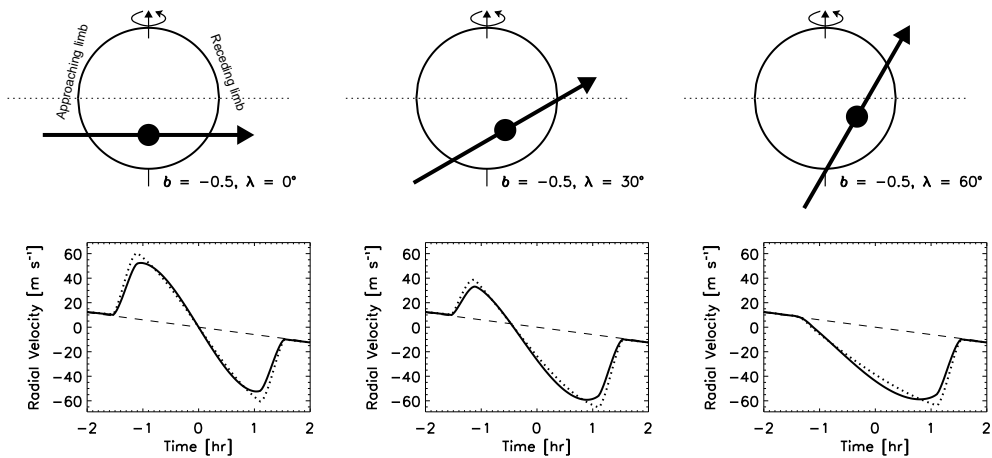}
% \vspace*{-1.0 cm}
 \caption{ The Rossiter-McLaughlin (RM) effect as an ``anomalous''
   Doppler shift.  Top: three possible transit geometries that produce
   identical light curves, but differ in spin-orbit alignment.
   Bottom: The corresponding radial-velocity signals.  Dotted lines
   are for the idealized case of no limb darkening; solid lines
   include limb darkening. }
   \label{fig7}
\end{center}
\end{figure}

\begin{table}
  \begin{center}
  \caption{Results for the projected spin-orbit angle, $\lambda$.}
    \label{tab2}
    \begin{tabular}{lcc}\hline
      {\bf System} & $\lambda$~[deg] & {\bf Refs.} \\ \hline
      HD~189733 & $-1.4\pm 1.1$ & 1 \\
      HD~209458 & $-4.4\pm 1.4$ & 2,3$^\star$ \\
      HAT-P-1   & $+3.6\pm 2.0$ & 4 \\
      CoRoT-Exo-2 & $+7.2\pm 4.5$ & 5 \\
      HD~17156 & $+9.4\pm 9.3$ & 6,7$^\star$ \\
      TrES-2 & $-9\pm 12$ & 8 \\
      HAT-P-2 & $+1\pm 13$ & 9$^\star$,10 \\
      HD~149026 & $-12\pm 15$ & 11 \\
      XO-3 & $+70\pm 15$ & 12 \\
      WASP-14 & $-14\pm 17$ & 13 \\
      TrES-1 & $+30\pm 21$ & 14 \\
      \hline
    \end{tabular}
  \end{center}
  \vspace{1mm}
  {\it References:}
  (1) Winn et al.~(2006).
  (2) Queloz et al.~(2000). (3) Winn et al.~(2005).
  (4) Johnson et al.~(2008).
  (5) Bouchy et al.~(2008).
  (6) Narita et al.~(2008). (7) Cochran et al.~(2008).
  (8) Winn et al.~(2008c).
  (9) Winn et al.~(2007d). (10) Loeillet et al.~(2008).
  (11) Wolf et al.~(2007).
  (12) Hebrard et al.~(2008).
  (13) Joshi et al.~(2008).
  (14) Narita et al.~(2007).
  Note: where more than one reference is listed, the value in
  column 2 is taken from the starred reference.
\end{table}

To date, RM observations have been reported for 11 different
exoplanetary systems, since the pioneering detections by Queloz et
al.~(2000) and Bundy \& Marcy~(2000). This field is at the transition
point between a mere handful of results to a large enough sample for
meaningful general conclusions to be drawn and for rare surprises to
emerge. The results are summarized in Table~2, which is organized in
order of measurement precision, with the most precise result listed
first. Figure~8 shows some more recent data for HD~189733 and HAT-P-2,
both of which display the antisymmetric waveform indicative of a
well-aligned spin and orbit (at least in projection). Both stars have
candidate stellar companions, naturally raising the possibility of
Kozai cycles, but in both cases the projected spin-orbit angle is
consistent with zero.

\begin{figure}[t]
% \vspace*{-2.0 cm}
\begin{center}
 \includegraphics[width=32pc]{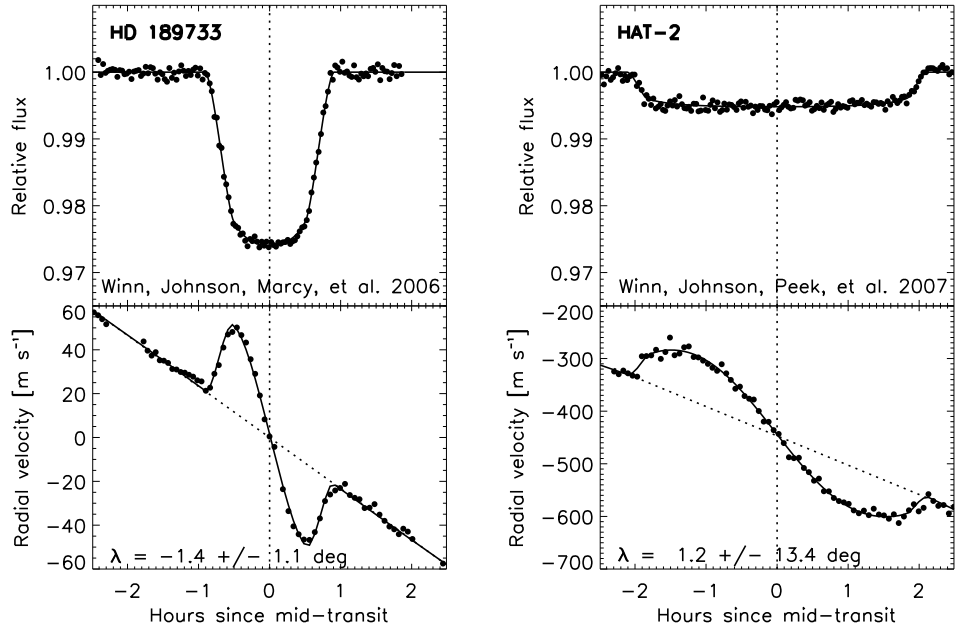}
% \vspace*{-1.0 cm}
 \caption{ Examples of RM data.  The top panels show transit
   photometry, and the bottom panels show the corresponding
   radial-velocity waveform.  The out-of-transit gradient in the
   radial velocity is due to the orbital motion of the star.  The
   in-transit ``blip'' is the RM effect. }
   \label{fig8}
\end{center}
\end{figure}

In almost all cases, the results are consistent with good alignment,
with measurement precisions ranging from $1^\circ$ to $21^\circ$.
There are two possible exceptions. The first is HD~17156, for which
Narita et al.~(2008) reported $\lambda = 62^\circ \pm 25^\circ$.
However, shortly after this meeting, Cochran et al.~(2008) presented
new data giving $9.4^\circ \pm 9.3^\circ$. The reason for the
apparently significant disagreement between the data sets is not yet
clear. The other possible exception is XO-3, for which H\'ebrard et
al.~(2008) found $\lambda = 70^\circ \pm 15^\circ$~deg, i.e., a
transverse RM effect in which only an anomalous blueshift was seen.
Those authors expressed some caution about the result due to possible
systematic errors.  Further observations of both of these systems are
warranted.

Broadly speaking, the results show that large spin-orbit misalignments
are fairly rare. The time is now ripe for a statistical analysis of
the ensemble results. This would not only overcome the sky-projection
limitation that is inherent in individual results, but could
ultimately place constraints on the fraction of systems that have
migrated via different channels. A possible confounding factor is
tidal coplanarization: even if the migration process resulted in a
large value of $\psi$, could the system have been eventually realigned
by star-planet tidal interactions? The simple and widely-used model of
tidal interactions, in which the equilibrium tidal bulge is shifted by
a constant time lag, suggests that the timescale for inclination
damping is much longer than the timescale for eccentricity damping,
and that tidal spin-orbit realignment can be neglected (Winn et
al.~2005). However, more theoretical work is needed to go beyond this
order-of-magnitude argument and define the scope for more complicated
tidal and evolutionary models to affect the interpretation of the RM
results. It is possible, for example, that tidal dynamics were more
important when the system was younger, or that the coplanarization
timescale is shorter than expected (Mazeh 2008).

\acknowledgements Special thanks are due to D.~Fabrycky for producing
Fig.~6, to A.~P\'al for providing the numerical results in Fig.~4
(which are more complete than the results I showed in my
presentation), and to K.~de~Kleer for help in producing Fig.~3. I have
enjoyed and benefited from collaborations with M.~Holman, J.~Johnson,
G.~Marcy, N.~Narita, Y.~Suto, D.~Fabrycky, E.~Turner, S.~Gaudi,
T.~Mazeh, J.~Carter, M.~Nikku, and others too numerous to mention, on
the topics presented here. I am grateful to the SOC for the
opportunity to speak at the Woodstock of transiting planets.

\end{document}